\documentclass[12pt,pdftex]{article}
\usepackage[dvips]{graphicx}
\usepackage{amsmath,amssymb,amsthm}
\usepackage{mathbbol,bbm}
\usepackage[final]{showkeys}
\usepackage{bm}
\usepackage{calc}

\begin{document}

\begin{center}
{\LARGE Thermodynamics of Markovian Open Quantum Systems with Application to Lasers}\\
R.~Alicki\\
International Centre for Theory of Quantum Technologies (ICTQT) \\
University of Gda\'nsk, Poland \\[6pt]
\end{center}

\begin{abstract}
G\"oran Lindblad  was one on the pioneers of what is called now Quantum Thermodynamics. From this vast and rapidly developing field we have selected a sample of results concerning quantum open systems described by Markovian Master Equations of the Lindblad (Gorini-Kossakowski-Sudarshan) type, which are applied to models of lasers. One can study their thermodynamics  using the properties of quantum relative entropy, also introduced by Lindblad.
\end{abstract}

\section{Introduction}

Quantum Thermodynamics (QT) is nowadays a vast field of research involving methods of mathematical and theoretical physics as well as various  experimental implementations \cite{QT-review}- \cite{KK}. From the very beginning QT is strongly associated with the theory of Quantum Open Systems (QOS)  whose aim is to describe the irreversible evolution of a quantum system interacting with an environment  \cite{Davies} -\cite{Huelga}.
Trace preserving completely positive maps $\Lambda$ was recognized in 1971 by Kraus \cite{Kraus}  as the proper candidate for a general irreversible dynamical map acting on the reduced density matrix of   QOS. In 1975 Lindblad proved \cite{Lindblad_ent} that the relative entropy defined for two density matrices as ($k_B \equiv 1$)
\begin{equation} 
S(\rho_1|\rho_2) = \mathrm{Tr}(\rho_1 \ln \rho_1 - \rho_1 \ln \rho_2)
\label{rel_entropy}
\end{equation}
satisfies monotonicity condition with respect to any dynamical map $\Lambda$
\begin{equation} 
S(\Lambda\rho_1|\Lambda\rho_2) \leq S(\rho_1|\rho_2) .
\label{rel_entropy}
\end{equation}
Treating the relative entropy as a kind of  ``distance'' between two quantum states one can say that any dynamical map reduces their distinguishability what seems to be the fundamental feature of irreversibility accompanied by inevitable information loss. The inequality \eqref{rel_entropy} finds numerous applications in QOS, QT and quantum information.

Another pillar of the theory of QOS is the general form of the Markovian Master Equation (MME) for the density matrix $\rho(t)$ of the open system satisfying complete positivity (CP) condition. This form was  presented in the year 1976 in two independent papers by Lindblad \cite{L} and Gorini--Kossakowski--Sudarshan \cite{GKS}  ($\hbar\equiv 1$) 
\begin{equation} 
\frac{d}{dt}{\rho}= -i[H , \rho] + \frac{1}{2}\sum_{j} ([V_j \rho, V_j^{\dagger}]+[V_j ,\rho V_j^{\dagger}]) \equiv\mathcal{L}\rho\, .
\label{GKLS}
\end{equation}
The Heisenberg form of the evolution equation valid for the observable $A(t)$ is also useful
\begin{equation} 
\frac{d}{dt}A= i[H , A] + \frac{1}{2}\sum_{j} ( V_j^{\dagger}[A ,V_j]+ [V_j^{\dagger} , A ] V_j^{\dagger}) \equiv\mathcal{L}^* A\, .
\label{GKLS_H}
\end{equation}
Combining both \eqref{rel_entropy} and the solution of \eqref{GKLS} written as  $\Lambda_t = e^{\mathcal{L}t}$ one obtains the Spohn inequality \cite{Spohn}
\begin{equation} 
\mathrm{Tr}\left(\mathcal{L}\rho (\ln \rho - \ln \rho_{st})\right) = \mathrm{Tr}\left(\rho \mathcal{L}^*(\ln \rho - \ln \rho_{st})\right)\leq 0, 
\label{Spohn}
\end{equation}
valid for an arbitrary  state $\rho$ and the stationary state $\rho_{st}$ of the evolution given by the GKLS generator $\mathcal{L}$. Adding to the game the consistent derivations of MME-s
based either on the weak coupling limit \cite{WCL} or low density limit \cite{LDL} one obtains the foundation of Thermodynamics of Markovian Open Quantum Systems. It is remarkable that this approach preceded the development of classical stochastic thermodynamics  \cite{Seifert} similarly to von Neumann's quantum probability which preceded Kolmogorov's axioms for classical probability theory.

The notion of Markovian dynamics can be extended to inhomogeneous in time evolutions governed by the MME of the form
\begin{equation} 
\frac{d}{dt}{\rho(t)} =\mathcal{L}(t)\rho(t)\, .
\label{GKLSn}
\end{equation}
where  $\mathcal{L}(t)$ is a time-dependent GKLS generator. The propagator solving eq. \eqref{GKLSn} can be written as  a limit of the composition of CP-maps
\begin{equation} 
\Lambda(t , s)  = \lim_{n\to\infty} e^{\mathcal{L}(u_n) (t-u_n)}\cdots  e^{\mathcal{L}(u_1) (u_2 - u_1)} e^{\mathcal{L}(s) (u_1 - s)}, \quad t\geq s.
\label{prop}
\end{equation}
For each component $\Lambda_k = e^{\mathcal{L}(u_{k-1}) (u_k - u_{k-1})}$  in \eqref{prop} one can construct a  model of the reservoir  $R_k$ with its own state $\rho_{R_k}$ and the joint unitary evolution $U_k$ such that
\begin{equation} 
\Lambda_k \rho = \mathrm{Tr}_{R_k}\left(U_k \rho\otimes\rho_{R_k} U^{\dagger}_k\right) .
\label{dilation}
\end{equation}
As noticed in \cite{Durban} this construction corresponds to the most general physical idea of memoryless (Markovian) model where the system is  coupled to a permanently ``refreshed'' reservoir such that the temporal state of the total system is a product state and the total temporal dynamics is unitary. Various definitions of Markovianity with the associated measures of non-Markovianity were developed  in the recent years by different authors (see for a review \cite{nonM}). Remarkably, the combination of the composition property $\Lambda(t , s) = \Lambda(t , u)  \Lambda(u , s) $ with the Lindblad monotonicity  \eqref{rel_entropy} implies non-increasing of the relative entropy distance between two solutions of the MME \eqref{GKLSn}. The increase of this distance observed for non-Markovian evolution can be interpreted in terms of the ``information back-flow from the reservoir''. It follows that the entropy production in its strict, local in time, version is always positive only for Markovian systems, hence for the non-Markovian ones the second law can be satisfied only in approximate, coarse-grained in time sense.

\section{Quantum Thermodynamics with additive environment}

The simplest scheme leading to  consistent description of open system thermodynamics can be summarized by the following MME \cite{SpohnLeb}, \cite{Alicki:79}
\begin{equation} 
\frac{d}{dt}{\rho}(t) = -i[H(t) , \rho] + \sum_k \mathcal{L}_k(t)\rho(t) .
\label{MMEadd}
\end{equation}
Here $H(t)$ is the physical (renormalized) Hamiltonian of the system with time dependence representing the deterministic external driving.   Each generator  $\mathcal{L}_k(t)$ accounts for the independent dissipative interaction with the $k$-th thermal reservoir at the inverse temperature $\beta_k$ (\emph{additive environment}) such that the temporal thermalization condition is satisfied 
\begin{equation} 
\mathcal{L}_k(t)\rho_{\beta_k}(t) = 0, \quad \rho_{\beta_k}(t) \equiv Z_k^{-1}(t) e^{-\beta_k H(t)}.
\label{therm}
\end{equation}
Such MME can be derived using weak coupling limit technique combined with the adiabatic approximation which are valid in the weak coupling regime and for time scale of the  $H(t)$ variations much longer then the time scale of the relevant temporal Bohr frequencies of the system.
\par 
Defining the heat current flowing to the system from the $k$-th bats as 
\begin{equation} 
J_k(t)  =\mathrm{Tr}\left[\rho(t)\mathcal{L}_k^*(t) H(t) \right] .
\label{heatcurrent}
\end{equation}
one can use the inequality \eqref{Spohn} applied to each bath separately  
to derive the Second Law in the form
\begin{equation} 
\frac{d}{dt}S(t) - \sum_k  \beta_k J_k(t)  \geq 0
\label{2Law}
\end{equation}
with the entropy $S(t) = -\mathrm{Tr}[\rho(t) \ln \rho(t)]$.
\par
With the definition of heat currents \eqref{heatcurrent} one obtains the First Law of thermodynamics 
\begin{equation} 
\frac{d}{dt}E(t) = \sum_k  J_k(t) - P(t), \quad  P(t) = -\mathrm{Tr}\left(\rho(t) \frac{d H(t)}{dt} \right) .
\label{MMEadd}
\end{equation}
where $E(t) =\mathrm{Tr}\left(\rho(t) H(t) \right)$ is the internal energy of the system and $P(t)$ is the power output absorbed by the external driving.
\par
Notice that, here,  the work reservoir which typically corresponds to a semi-classical meso- or macroscopic device extracting (supplying) a useful work from (to) the engine or refrigerator is modeled by a  deterministic classical system hidden in the time-dependent part of the Hamiltonian. Therefore,  to obtain a fully quantum theory of autonomous engines and refrigerators we have to extend the formalism including quantum models of pistons or turbines. 
\par
Finally, one should mention that the additive reservoir model with external driving can be derived also for periodic in time driving with frequency higher or comparable to relevant Bohr frequencies and even for non-equilibrium but stationary baths, characterized by ``local'', frequency dependent temperatures as long as the system-baths couplings can be considered as weak \cite{localT}.

\section{Quantum Thermodynamics with multiplicative environment}
Now we discuss   a quantum open system which can be called chemical engine, with the quantum piston being a quantum harmonic oscillator (a simplified version of the model used in \cite{AlGKLS}). In contrast to the model from the previous Section, here the piston is a Markovian open system and the other degrees of freedom usually attributed to working fluid are included in two chemical baths. Bath A corresponds to reactants and bath B to reaction products while the excitations of the piston are called photons, because laser theory is a natural application domain for such a model. The symbolic chemical reaction can be written as
\begin{equation}
A \rightleftharpoons B + X.
\label{reaction}
\end{equation}
where $X$ denote emitted or absorbed photon. The Hamiltonian of the total system is given by 
\begin{equation}
\hat{H}_{tot}= \omega_0 a^{\dagger} a + H_A + H_B  + H_{int}\, ,
\label{hamtot}
\end{equation}
where $H_j , j= A, B,$ are Hamiltonians of chemical baths and the interaction Hamiltonian is given by
\begin{equation}
H_{int}= a\otimes R^{\dagger}  + {a}^{\dagger}\otimes {R}\,  .
\label{hamint}
\end{equation}
Here,  $a^{\dagger}$  creates a single photon and $R$ annihilates the molecules $A$  and creates the molecule $B$ while their hermitian adjoints correspond to time-reversed processes. Using the Fock space model for many-body systems we can write the operator $R$ as  sum of \emph{products} of the annihilation operator for the A-molecule and the creation operator for the B-molecule what explains the name \emph{multiplicative environment}.
\par
The thermodynamical equilibrium state of the environment is a joint great canonical ensemble for both types of molecules written as
\begin{equation}
{\rho}_R = Z^{-1} \exp\Bigl\{-\beta \sum_{j=A,B}\bigl({H}_j - \mu_j {N}_j\bigr)\Bigr\}, \quad [{N}_j , {H}_{j'} ] = 0 ,.
\label{chemeq}
\end{equation}
with the operator ${N}_j$ counting the number of $j$-type molecules and $\mu_j$ being the corresponding chemical potential , $j = A,B$.
\par
The details of the structure of Hilbert spaces describing molecules are not relevant, only the following relation describing the energy conservation is crucial 
\begin{equation}
[\mu_A {N}_A + \mu_B {N}_B ,  {R}] = (\mu_B - \mu_A ) {R} .
\label{chemcondition}
\end{equation}
The interaction Hamiltonian \eqref{hamint} can be seen as a coarse-grained  description of a complicated reversible quantum process with the final effect \eqref{reaction}.
\par
In the next step one can derive, using standard methods combining Born and secular approximations the GKLS equation for the harmonic oscillator density matrix
\begin{equation}
\frac{d}{dt} {\rho}  = -i\omega [{a}^{\dagger}{a}, {\rho}] +\frac{\gamma_\downarrow }{2}\bigl([{a}, {\rho}{a}^{\dagger}] + [{a} {\rho}, {a}^{\dagger}]\bigr)+ \frac{\gamma_\uparrow}{2}\bigl([{a}^{\dagger}, {\rho} {a}] + [{a} {\rho}, {a}^{\dagger}]\bigr).
\label{MEosc}
\end{equation}
with the standard expressions for damping and pumping rates
\begin{equation}
 \gamma_{\downarrow}  =\int_{-\infty}^{\infty}e^{i\omega t}\,\mathrm{Tr}\bigl({\rho}_R {R}(t){R}^ {\dagger}\bigr) dt ,\quad \gamma_{\uparrow}  =\int_{-\infty}^{\infty}e^{i\omega t}\,\mathrm{Tr}\bigl({\rho}_R {R}^{\dagger}(t) {R}\bigr) dt .
\label{relaxation}
\end{equation}
and renormalized photon frequency $\omega$.
\par
Due to \eqref{chemeq} \eqref{chemcondition} they satisfy the detailed balance condition for chemical baths at isothermal conditions 
\begin{equation}
\frac{\gamma_{\uparrow}}{\gamma_{\downarrow}}   = \exp\{-\beta{\Delta G}\}\,   
\label{dbchem}
\end{equation}
with $\Delta G$ interpreted as the  Gibbs free energy released in the reaction,
\begin{equation}
\Delta G = \omega + \mu_B - \mu_A .
\label{gibbsfree}
\end{equation}
The reaction \eqref{reaction}  from left to right is spontaneous under the following condition  
\begin{equation}
\Delta G < 0 \Rightarrow \gamma_\uparrow > \gamma_\downarrow
\label{amplification}
\end{equation}
Notice that in this case the energy is constantly pumped to the piston as shown by the formula
\begin{equation}
E(t)= \omega \mathrm{Tr} \left({\rho}(t) {a}^{\dagger}{a}\right) = e^{(\gamma_\uparrow - \gamma_\downarrow)t} W(0) + \bigl[e^{(\gamma_\uparrow - \gamma_\downarrow)t} -1\bigr] \frac{\omega\gamma_\uparrow }{\gamma_\uparrow - \gamma_\downarrow} , 
\label{energy}
\end{equation}
and therefore the stationary state for the eq.\eqref{MEosc} does not exist. However, using the commutation relation
\begin{equation}
e^{z a^{\dagger}a} a = e^{-z}ae^{ z a^{\dagger}a}, \quad z \in \mathbf{C}
\label{commutation}
\end{equation}
one can check that the positive unbounded operator 
\begin{equation}
\rho_{st} = e^{-\beta( \omega +\mu_B -\mu_A)  a^{\dagger}a}
\label{negtemp}
\end{equation}
formally, satisfies the relation $\mathcal{L}\rho_{st} = 0$ with $\mathcal{L}$ given by \eqref{MEosc} - \eqref{amplification}.
One can notice that the inequality \eqref{Spohn} does not depend on the stationary state  normalization. Therefore, one can use it also for our unbounded ``stationary state'' \eqref{negtemp} to obtain the following form of the Second Law for our model
\begin{equation} 
\frac{d}{dt}S(t) - \beta J(t)  \geq 0
\label{2Law}
\end{equation}
with the heat current flowing from the isothermal environment defined as
\begin{equation} 
J(t)  =(\omega +\mu_B - \mu_A)\frac{d}{dt}N(t) .
\label{heatcurrentmulti}
\end{equation}
where
\begin{equation} 
N(t)  =\mathrm{Tr}[\rho(t) a^{\dagger}a ], \quad \frac{d}{dt}N(t)=\mathrm{Tr}\left[\rho(t)\mathcal{L}^* (a^{\dagger}a) \right] \equiv j.
\label{particleN}
\end{equation}
Taking into account \eqref{reaction} and \eqref{chemcondition} which describe molecule/photon balance we can define molecular currents from the chemical reservoirs as
\begin{equation} 
\frac{d}{dt}N_A (t)\equiv j_A = -j , \quad  \frac{d}{dt}N_B (t)\equiv j_B = j
\label{particlecurrent}
\end{equation}
what allows to write the First Law in a form typical for chemical systems
\begin{equation} 
\frac{d}{dt}E(t) = J -\mu_A j_A - \mu_B j_B.
\label{1Law}
\end{equation}
However, it is still not obvious whether the whole internal energy of the oscillator $E(t)$ can be interpreted as work extracted by the piston from the chemical baths. To illustrate this problem, consider the phase-space picture of the evolution of  initial coherent state $\rho(0) =|\alpha_0\rangle\langle\alpha_0|$, as shown on Fig.2. The temporal Gaussian state with the center following the classical trajectory  of exponentially expanding spiral exhibits also exponentially increasing width  indicating the corresponding entropy increase. On the other hand work is the energy supplied  to the piston at the negligible entropy transfer.

\begin{figure} [t] 
\begin{center}
	\includegraphics[width=0.60 \textwidth]{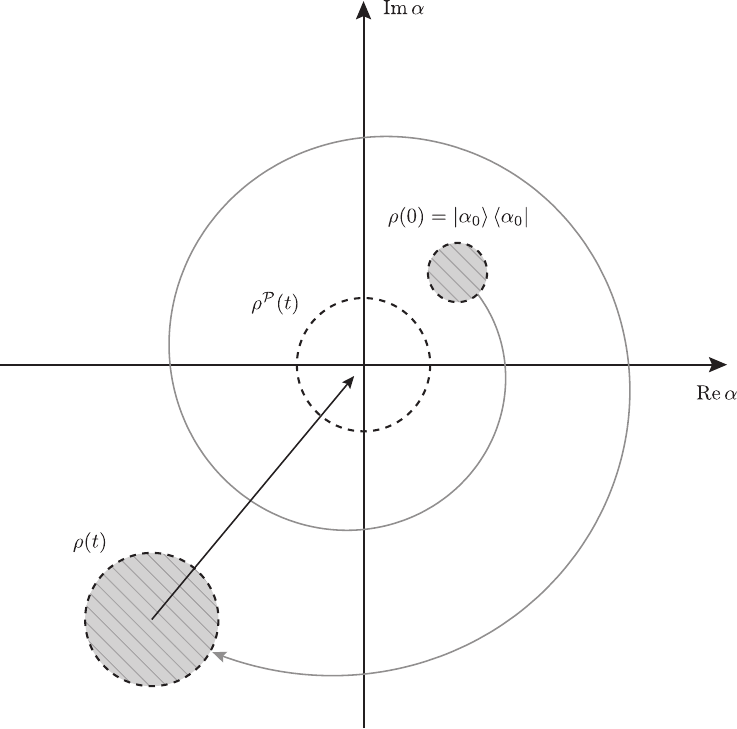}
\end{center}
\caption{\small  \textbf{Phase-space picture of harmonic oscillator evolution.} The initial coherent state $\rho(0) =|\alpha_0\rangle\langle\alpha_0|$ evolves into the mixed Gaussian state $\rho(t)$ which shifted to the origin yields the passive state $\rho^{\mathcal{P}}(t)$ \cite{AlGKLS}.}
\end{figure}

To eliminate the useless energy hidden in the piston's fluctuations we can identify  the accessible work stored in its state $\rho$ with the ergotropy \cite{ergo}
\begin{equation}
W_e =  \mathrm{Tr} ({\rho}{H}) -\mathrm{Tr} ({\rho}^{\mathcal{P}}{H}).
\label{ergotropy}
\end{equation}
where $\rho^{\mathcal{P}}$ is a passive state with respect to the Hamiltonian $H= \omega {a}^{\dagger}{a}$ unitarily equivalent to ${\rho}$ \footnote{Passive state with respect to the given Hamiltonian is diagonal
in the energy basis with eigenvalues ordered in a decreasing order with the energy \cite{Pusz}}.

\par
Another definition of accessible work, used e.g. in \cite{Seah}, is based on the semiclassical expression for energy 
\begin{equation}
W_{sc} =  \omega |\mathrm{Tr} ({\rho}a) |^2 \leq\mathrm{Tr} ({\rho}{H})
\label{semiclassocal}
\end{equation}
which separates the energy of deterministic component of the piston's motion from thermal and quantum fluctuations.
\par
The obtained above piston's oscillations with exponential increase of the amplitude can describe the initial stage of engine's dynamics, like for example the onset  of laser action.  For real systems we are rather interested in the stationary regime. To achieve this an external load must be  attached to the piston stabilizing its dynamics for long times. The proper construction of a model of stationary  work extraction in quantum engines is still not completely solved (see e.g. an interesting proposal, called \emph{quantum flywheel}, presented in \cite{Levy:2016}). A particular example of external load model is discussed in the following section.

\section{Laser as an engine in the stationary regime}

There exist many examples of  phenomenological constructions of MME describing laser action in the simplest case of  a single mode laser. We are interested in a particular class of MME being the extension of the equation \eqref{MEosc} in such a way that stationary solutions describing coherent light exist above the threshold (some examples are discussed in \cite{Hen}). To simplify the formulas we use the notation
\begin{equation} 
\mathcal{D}[V] \rho \equiv \frac{1}{2} ([V \rho, V^{\dagger}]+[V,\rho V^{\dagger}]) .
\label{Lindblad}
\end{equation}
Then one can discuss the properties of MME for the harmonic oscillator of the form
\begin{equation}
\frac{d}{dt} {\rho}  = -i\omega [{a}^{\dagger}{a}, {\rho}] +\mathcal{D}[a g_{\downarrow}(a^{\dagger}a)] \rho +\mathcal{D}[a^{\dagger} g_{\uparrow}(a^{\dagger}a)]  \rho  .
\label{MEosc1}
\end{equation}
determined by two functions $ g_{\downarrow}(x)$, $g_{\uparrow}(x)$. Notice that for $ g_{\downarrow}(x )= \sqrt{\gamma_{\downarrow}}$, $g_{\uparrow}(x)=\sqrt{\gamma_{\uparrow}}$ we recover \eqref{MEosc}. The diagonal matrix elements $p_n \equiv \langle n|\rho|n\rangle$ describing probability of detecting $n$ photons evolve independently of the off-diagonal ones according to the \emph{birth and death process}
\begin{equation}
\frac{d}{dt} p_n(t)= \Gamma_\downarrow [n +1] p_{n+1}(t) + \Gamma_\uparrow [n-1] p_{n-1}(t) 
 - \left( \Gamma_\downarrow [n]+\Gamma_\uparrow [n] \right) p_{n}(t)~.
\label{eq:birth}
\end{equation}
with the  rates 
\begin{equation}
 \Gamma_\uparrow [n] = (n+1) |g_{\uparrow}(n) |^2 , \quad  \Gamma_\downarrow [n] =  n |g_{\downarrow}(n)|^2
\label{eq:birthrates}
\end{equation}
The diagonal stationary state  for the MME \eqref{MEosc1}, determined by the probability distribution $\{\bar{p}_n\}$,  exists under the sufficient condition
\begin{equation}
\lim_{n\to\infty}\frac{ \Gamma_\uparrow [n-1] }{\Gamma_\downarrow [n] } < 1
\label{eq:stationarycond}
\end{equation}
and is given by the expression
\begin{equation}
\bar{p}_n = \bar{p}_0 \prod_{k=1}^n\frac{ \Gamma_\uparrow [k-1] }{\Gamma_\downarrow [k] }, \quad  n > 0 ,
\label{eq:stationarystate}
\end{equation}
where $\bar{p}_0$ is determined by the normalization condition.
\par
For linear pumping and damping
\begin{equation}
 \Gamma_\uparrow [n] = \gamma_{\uparrow}(n+1) , \quad  \Gamma_\downarrow [n] =  \gamma_{\downarrow} n
\label{eq:birthrates1}
\end{equation}
what according to \eqref{eq:stationarycond} and \eqref{eq:stationarystate} implies that below the threshold ($\gamma_{\uparrow} <\gamma_{\downarrow}$)  the thermal state 
\begin{equation}
\bar{p}_n = \left(1 -e ^{-\beta\omega}\right) e^{-\beta\omega n} ,\quad  \frac{\gamma_{\uparrow}}{\gamma_{\downarrow}}= e^{-\beta\omega}
\label{eq:statstate1}
\end{equation}
is stationary, while  above  the threshold ($\gamma_{\uparrow} >\gamma_{\downarrow}$)  laser action begins with unlimited photon number grow. In the later case one can say that the mode is coupled to the bath at the \emph{negative} inverse temperature.
\par
To obtain the  realistic description of  a laser acting above the threshold one has to introduce nonlinearity in pumping or/and damping processes. The two simplest examples are given by the two following definitions of model parameters expressed in the standard notation of Einstein coefficients $A$ and $B$ and the new parameter, $0 <C << 1$, characterizing nonlinearity
\begin{equation}
 \Gamma_\uparrow [n] = \frac{A(n+1)}{1 + C(n+1)} , \quad  \Gamma_\downarrow [n] =  B n
\label{eq:birthratesA}
\end{equation}
\begin{equation}
 \Gamma_\uparrow [n] = A (n+1) , \quad  \Gamma_\downarrow [n] =  B (n + C n^2)
\label{eq:birthratesB}
\end{equation}
They correspond to saturation of pumping or damping process, respectively, but both lead to the same ratio
\begin{equation}
\frac{ \Gamma_\uparrow [n-1] }{\Gamma_\downarrow [n] } = \frac{A}{B} \frac{1}{1+ C n}
\label{eq:stationarycond1}
\end{equation}
which due to the condition \eqref{eq:stationarycond} assures  the existence of a stationary state for all  $A, B, C > 0$. One can easily check that for $ A/B <<1$ this stationary state is close to the thermal one \eqref{eq:statstate1} while for $A/B >> 1$ we obtain approximately the Poisson distribution
\begin{equation}
\bar{p}_k \simeq e^{-\bar{n}}\frac{ \bar{n}^k}{k!}, \quad  \bar{n} = \frac{A}{BC}
\label{eq:stationarystate}
\end{equation}
corresponding to the phase-averaged coherent state of the mode in agreement with the laser phenomenology.
\par

In the following a  different proposal inspired by the previous paper dealing with classical stochasic models \cite{engines} is discussed. The laser is still coupled to the chemical baths discussed in Section 3  but the \emph{load} is modeled by a non-linear macroscopic friction proportional to the new parameter $\delta >0$. The full MME reads
\begin{equation}
\frac{d}{dt} {\rho}  = -i\omega [{a}^{\dagger}{a}, {\rho}] +\mathcal{L}_{\mathrm{bath}}\rho +\mathcal{L}_{\mathrm{load}}\rho 
\label{MEosc2}
\end{equation}
with
\begin{equation}
\mathcal{L}_{\mathrm{bath}} =\gamma_{\downarrow}\mathcal{D}[a] +\gamma_{\uparrow}\mathcal{D}[a^{\dagger}], \quad  \mathcal{L}_{\mathrm{load}}= \delta\mathcal{D}[a \sqrt{a^{\dagger}a}]  .
\label{MEosc2}
\end{equation}
The diagonal elements of $\rho(t)$ evolve according to the birth and death process \eqref{eq:birth} with the rates
\begin{equation}
 \Gamma_\uparrow [n] = \gamma_{\uparrow}(n+1) , \quad  \Gamma_\downarrow [n] =  \gamma_{\downarrow} n + \delta n^2
\label{eq:birthratesC}
\end{equation}
Obviously, the dynamics of diagonal matrix elements and hence also the stationary state $\bar{\rho} = \sum_n \bar{p}_n |n\rangle\langle n|$ is the same as for the previous model \eqref{eq:birthratesB} ($A = \gamma_{\uparrow}, B =\gamma_{\downarrow}, C = \delta/\gamma_{\downarrow}$). However, the different form of the MME \eqref{MEosc2} with explicitly defined  action of the load allows for a consistent thermodynamic interpretation.
\par
The energy balance in the stationary state can be interpreted as the first law of thermodynamics and formulated in terms of the definitions \eqref{heatcurrentmulti}, \eqref{particlecurrent} with the modified form of the  molecular currents
\begin{equation} 
j\equiv \mathrm{Tr}\left[\rho(t)\mathcal{L}_{\mathrm{bath}}^* (a^{\dagger}a) \right] = j_B(t) = -j_A(t)
\label{particleN1}
\end{equation}
and the following definition of the work output transferred to the load
\begin{equation} 
P(t)=-\omega \mathrm{Tr}\left[\rho(t)\mathcal{L}_{\mathrm{load}}^* (a^{\dagger}a) \right] = \omega\delta\mathrm{Tr}\left[\rho(t) (a^{\dagger}a)^2 \right]  .
\label{load}
\end{equation}
The first law of thermodynamics  reads
\begin{equation} 
\frac{dE}{dt}= J - \mu_A j_A - \mu_B j_B - P .
\label{Ilaw _stat}
\end{equation}
and the second law takes form
\begin{equation} 
\frac{d}{dt}S(t) - \beta J(t)  +  \delta \dot{S}(t) \geq 0 ,
\label{2Law1}
\end{equation}
where the \emph{residual entropy production} is given by
\begin{equation} 
 \delta \dot{S}(t) =- \mathrm{Tr}\left[\rho(t)\mathcal{L}_{\mathrm{load}}^* \ln{\rho}(t) \right] .
\label{2Law2}
\end{equation}
To prove \eqref{2Law1},\eqref{2Law2} one uses again the Spohn inequality \eqref{Spohn} applied to the not normalized state  \eqref{negtemp} which is a stationary state for  $\mathcal{L}_{\mathrm{bath}}$. The inequality \eqref{2Law1} agrees with the phenomenological formulation of the second law only if the  residual entropy production \eqref{2Law2} can be neglected in the entropy balance \eqref{2Law1}.
\par
To check the consistency of the above results one can compute the  output power and the residual entropy production for the stationary regime and under the condition $\delta << \gamma_{\downarrow},\gamma_{\uparrow}$ which implies that the stationary state can be approximated by the  the phase-averaged coherent state  (see eq.\eqref{eq:stationarystate})
\begin{equation}
\bar{\rho}=  e^{-\bar{n}}\sum_{k=0}^{\infty}\frac{ \bar{n}^k}{k!}|k\rangle\langle k|, \quad  \bar{n} = \frac{\gamma_{\uparrow}}{\delta} .
\label{eq:stationarystate1}
\end{equation}
Then the approximate formula for the stationary output power reads
\begin{equation} 
\bar{P}\simeq \omega\delta \bar{n}^2 \simeq \frac{\omega \gamma^2_{\uparrow}}{\delta}.
\label{load1}
\end{equation}
To estimate $\delta \dot{S}$ in the diagonal state
\begin{equation}
{\rho} = \sum_{k=0}^{\infty} p_k |k\rangle\langle k|, 
\label{eq:diag}
\end{equation}
one inserts \eqref{eq:diag} into \eqref{2Law2} and uses \eqref{eq:birth} and \eqref{eq:birthratesC} to obtain
\begin{equation} 
 \delta \dot{S}(t) = \delta \sum_{k=0}^{\infty} \left[ k^2 p_k - (k+1)^2 p_{k+1}\right] \ln p_k .
\label{eq:residual}
\end{equation}
Assuming that $p_k$ is a smooth function of $k$ concentrated on the values of $k >> 1$ and using a discrete ``derivative''  $(\nabla f)_k = f_{k+1} - f_k $, and the ``integration by parts'' one can use the following approximations 
\begin{equation} 
 \delta \dot{S}(t) \simeq- \delta \sum_{k=0}^{\infty}\nabla(k^2 p_k) \ln p_k \simeq\delta \sum_{k=0}^{\infty}k^2 p_k\frac{ \nabla p_k}{p_k}\simeq - 2\delta \langle n\rangle,
\label{eq:residual1}
\end{equation}
where $\langle n\rangle = \sum_{k=0}^{\infty}k p_k $.
\par
Remembering that according to \eqref{eq:stationarystate1} the stationary photon number scales like $1/\delta$ then  $\delta \dot{S}$ remains roughly constant, independent of $ \langle n\rangle$, while the additive parameters of the laser radiation are proportional to $ \langle n\rangle >> 1$. This justifies neglecting  of $\delta \dot{S}$ in the entropy-energy balance.
\par
It is interesting that the similar residual entropy production appeared in the analysis of classical stochastic models presented in \cite{engines}. This mysterious contribution to entropy balance which can be neglected in the macroscopic limit was attributed to the oversimplified model of the load in terms of phenomenological friction force. In real systems work extraction contains  also the \emph{rectification process} involving additional degrees of freedom which contribute to the entropy balance.

\section{Concluding remarks}

Two fundamental contributions of G\"oran  Lindblad to the Quantum Theory of Open Systems and Quantum Thermodynamics - monotonicity of relative entropy with respect to dynamical maps and the form of Quantum Markovian Master Equations allowed to develop the mathematically and physically consistent formalism  describing irreversible processes in quantum domain. These results provide  clear guiding principles both, for derivations from first  principles and phenomenological constructions of the underlying dynamical equations.
Although this formalism became popular in the field of quantum optics and quantum information its potential applications are much broader ranging from solid state physics, chemistry  and biophysics to astrophysics and cosmology \cite{AJ:2018} - \cite{ABJ:2023}.

\end{document}